# Heat capacity of p-H$_2$-p-D$_2$-Ne solid solution: Effect of (p-D$_2$)Ne clusters

M.I. Bagatskii, I.Ya. Minchina, V.M. Bagatskii

*Verkin Institute for Low Temperature Physics and Engineering, National Academy of Sciences of Ukraine, 47, Lenin Ave., 61103, Kharkov, Ukraine*

e-mail: bagatskii@ilt.kharkov.ua

*The heat capacity of a solid solution of 1% p-D$_2$ and 0.25% Ne in p-H$_2$ has been investigated in the interval $\Delta T = 0.5 - 4$ K. An excess heat capacity $\Delta C_{Ne}$ of this solution exceeding the heat capacity of the 1% p-D$_2$ in p-H$_2$ solution has been detected and analyzed. It is found that below 2 K the dominant contribution to the heat capacity $\Delta C_{Ne}$ is made by the rotation of the p-D$_2$ molecules in the (p-D$_2$)Ne - type clusters. The number of (p-D$_2$)Ne clusters in the solid sample is strongly dependent on the conditions of preparation. The splitting of the J = 1 level of the p-D$_2$ molecules in the (p-D$_2$)Ne clusters $\Delta = 3.2$ K is consistent with the theoretical estimate Kokshenev.*

## INTRODUCTION

Heavy impurities in quantum crystals of hydrogen isotopes affect the phonon spectrum of the crystal and disturb zero (quantum) vibrations of the lattice and rotation of the molecules. Local changes in the lattice structure and the formation of new quantum objects (molecular clusters and complexes [1-8]) are also possible in the vicinity of heavy impurities, which can produce considerable changes in the physical properties of crystals. These phenomena have recently a focus of intensive investigations [7, 8].

The excess heat capacity ($\Delta C_{Ne}$) of the solution of 2.5% o-$H_2$ and $x$Ne ($x$ = 0.5%, 2%) in solid p-$H_2$ caused by the heavy quasi-isotopic Ne impurity introduced into the solid 2.5 % p-$D_2$ – p-$H_2$ solution was first observed in [1] at T = 2-6 K. Near T = 2 K $\Delta C_{Ne}$ is an order of magnitude over the results calculated in the harmonic approximation for $\Delta C_{L, Ne}$ caused by the heavy quasi-isotopic Ne impurity changing the phonon spectrum of the crystal.

A theory was put forward in [2] to explain this anomaly. Along with the anomalous heat capacity of solid $H_2$–Ne solutions attendant upon the change in the phonon spectrum of crystal, there is an anomaly in the rotational component of $C_{R, Ne}$ which is due to the contribution from the rotational degrees of freedom of the $H_2$ molecules in the lowest state with the rotational quantum number $J$ = 1 (o-$H_2$). The strong perturbation of zero (quantum) lattice vibrations by the Ne atoms in the (o-$H_2$)Ne – type clusters disturbs the local symmetry of the crystal field [2] causing the $J$ = 1 level of o-$H_2$ molecules to split into two levels with the degeneracies $g_0$ = 1 and $g_1$ = 2 (the upper level). (see Fig.1). When the (o-$H_2$)Ne cluster is formed, the energy of the subsystem decreases by $2\Delta/3$ ($\Delta$ is the splitting value). $C_{R, Ne}(T)$ exhibits a Schottky – type anomaly. The equilibrium contents of o-$H_2$o-$H_2$ and (o-$H_2$)Ne clusters in this system depends on temperature. As the temperature changes due to quantum diffusion of the angular momentum of the o-$H_2$ molecules, the number of clusters varies with time (configurational relaxation). The value of heat capacity is therefore dependent on the time $t_m$ of the (single) heat capacity measurement and the temperature prehistory. In [1] the heat capacity was measured above the temperature of the $C_{R, Ne}(T)$ maximum in the Schottky curve. The splitting 2.5 K < $\Delta$ < 5 K of the $J$ = 1 level of the o-$H_2$ molecules in the neighborhood of the Ne impurity was roughly estimated [2] from the analysis the results obtained in [1].

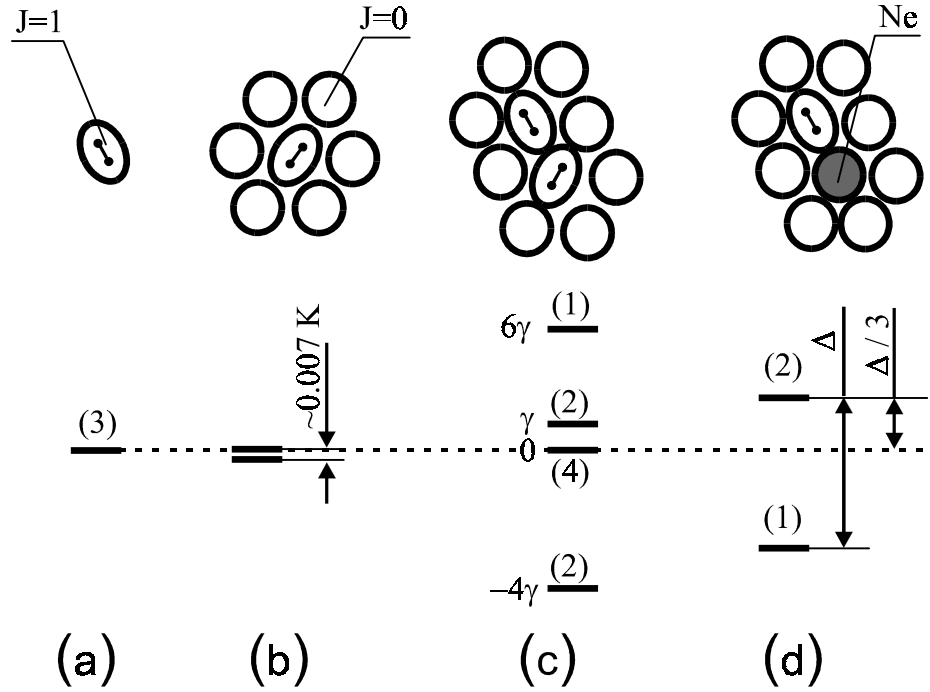

Fig. 1. Schematical arrangement of lower energy levels of o-$H_2$ (p-$D_2$) ($J = 1$) molecules as a function of molecular surroundings (level degeneracy is indicated in brackets): a) free molecule; b) twelve p-$H_2$ molecules ($J=0$ of the first coordination sphere of an hcp lattice (only six molecules are shown) [15]; c) two nearest neighboring o-$H_2$ molecules (cluster o-$H_2$o-$H_2$, $\gamma = 0.83$ K), p-$D_2$ (cluster p-$D_2$p-$D_2$, $\gamma = 0.95$ K) surrounded by the nearest neighboring p-$H_2$ molecules of the hcp lattice (eight molecules are shown) [15]; d) the nearest neighboring Ne atom and p-$D_2$ molecule ((p-$D_2$)Ne cluster) surrounded by the nearest neighboring p-$H_2$ molecules of the hcp lattice (eight molecules are shown) [2].

This study is concentrated on the contribution of the rotational motion of the p-$D_2$ molecules to the heat capacity of the solid 1% p-$D_2$ - p-$H_2$ solution doped with 0.25% Ne in the interval $\Delta T = 0.5 - 4$ K. The choice of impurity concentration and the temperature interval was dictated by the following considerations. Firstly, quantum diffusion of the p-$D_2$ molecules is impossible in the p-$H_2$ lattice [9] and conversion of the p-$D_2$ molecules during the experiment is negligible. Secondly, with the splitting $\Delta > 2$ K of the $J=1$ level of the p-$D_2$ molecules in the neighborhood of the Ne impurity the temperature of the maximum in the Schottky curves enters the temperature region of this investigation [2]. Thirdly, earlier we investigated the heat capacity of the solution of 1%p-$D_2$ in solid p-$H_2$ using the same calorimeter [9]. This permits us to separate accurately the excess heat capacity $\Delta C_{Ne}$ caused by 0.25% Ne introduced into the solid 1% p-$D_2$ – p-$H_2$ solution.

EXPERIMENT

The heat capacity of the solid solution of 0.94 mole % p-$D_2$ and 0.06 mole % o-$D_2$ in parahydrogen (below referred to as 1% p-$D_2$ in p-$H_2$) doped with 0.52% Ne has been measured using an adiabatic calorimeter [11] in the interval $\Delta T = 0.5 – 4$ K. The heat capacity of the solution 1% p-$D_2$ and 0.25% Ne in solid p-$H_2$ was measured using an adiabatic calorimeter [10]. The gas compositions were $H_2$ – 99.99% (the isotope – 99.985%, HD – 0.015%); $D_2$ – 99.99% (the isotope – 99.9%, HD – 0.1%); Ne- 99.99%. The starting orto-para composition of hydrogen $\approx 1\times10^{-2}$ % o-$H_2$ was obtained by keeping hydrogen in catalytic Fe(OH)$_3$ for 24 h at a constant temperature (the triple point of $H_2$). p-$D_2$ was obtained in an adsorption column by the technique described in [12]. The p-$D_2$ concentration (94%) in deuterium was estimated from the thermal conductivity of $D_2$ gas at nitrogen temperatures using an analyzer which we made and calibrated following the configuration in [13]. Four measurement series were performed. Series 1 was made on a sample prepared in the calorimetric vessel by condensing the gas mixture to the solid phase at $T\approx9.5$ K. The other series were made on solid samples prepared by crystallization from the liquid phase. After each series of measurement, the sample was melted, kept in the liquid state during a period of $t_i$ at temperature $T_i$, crystallized and cooled. Then the next run of measurement was performed.

Table 1. Time $t_i$ during which the sample was kept in the liquid phase near $T_i$

before its crystallization and the subsequent series of heat capacity measurement.

| series | 2 | 3 | 4 |
|---|---|---|---|
| $t_i$, min | 40 | 90 | 120 |
| $T_i$, K | 14.5 | 16 | 18 |

The heat capacities measured at $T \leq 4$ K are independent of the temperature prehistory of the sample. The measurement error was $\pm 6\%$ at 0.5 K, $\pm 2\%$ at 1K and $\pm 1\%$ at $T > 2$ K.

## RESULTS AND DESCUSSION

The experimental results on heat capacity per mole of the solution 1%p-$D_2$ and 0.25%Ne in solid p-$H_2$ can be written as

$$C = C_1 + \Delta C_{Ne} = C_1 + \Delta C_{L, Ne} + C_{R, Ne}. \tag{1}$$

Here $C_1$ is the heat capacity of the solid 1%p-$D_2$ – p-$H_2$ solution [9], $\Delta C_{Ne}$ is the excess heat capacity of the solution 1%p-$D_2$ and 0.25%Ne in solid p-$H_2$ over the heat capacity of the solution 1%p-$D_2$ in solid p-$H_2$. We assume that $\Delta C_{Ne} = \Delta C_{L, Ne} + C_{R, Ne}$, where $\Delta C_{L, Ne}$ is the increment in the heat capacity of the lattice produced by the quasi-local frequencies in the phonon spectrum of hydrogen, which appear when the heavy quasi-isotopic Ne impurity is introduced into the lattice of p-$H_2$; $C_{R, Ne}$ is the rotational heat capacity of the p-$D_2$ molecules caused by the 0.25% Ne impurity introduced into the p-$H_2$ - 1% p-$D_2$ solution. $\Delta C_{L, Ne}$ was calculated in the harmonic approximation using the technique developed by Peresada et al [13].

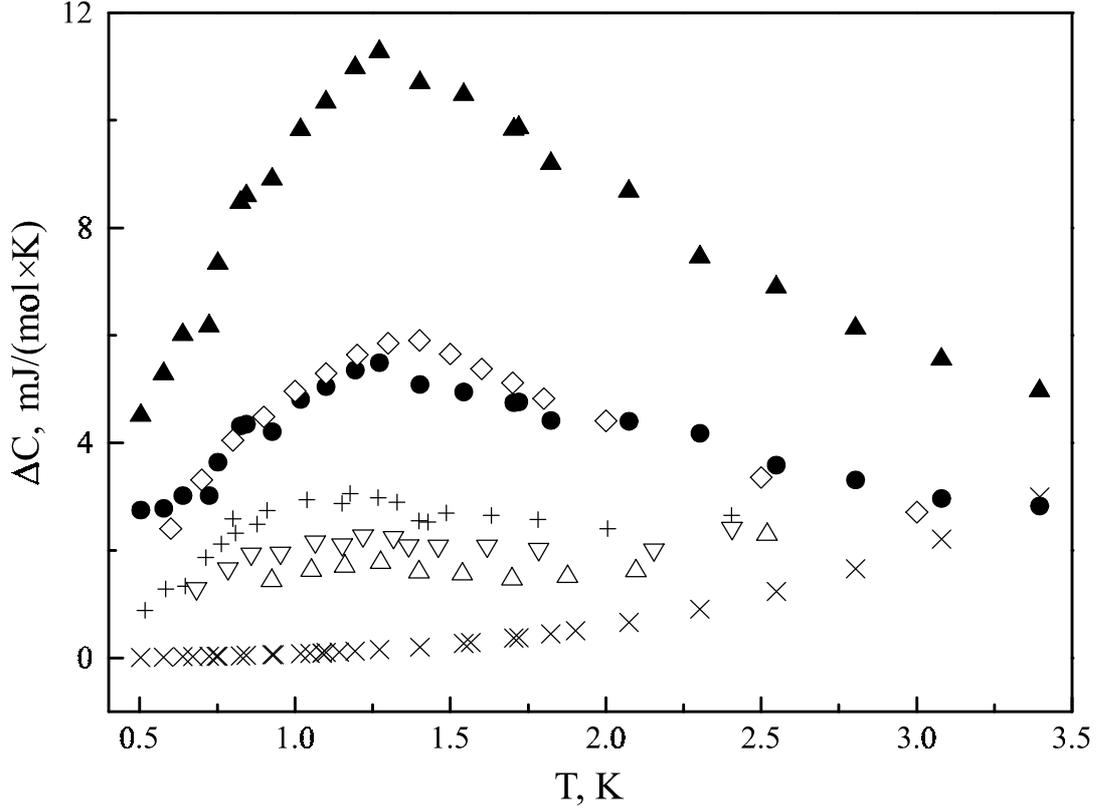

Fig. 2. Temperature dependences of excess heat capacities: ▲ - solution 1% p-D$_2$ and 0.25% Ne in solid p-H$_2$ over that of pure p-H$_2$ (series 1); ●, +, ▽, △ - solutions 1% p-D$_2$ and 0.25% Ne in solid p-H$_2$ over that of solutions 1% p-D$_2$ in solid p-H$_2$ (series 1-4, respectively); ◇ - solutions 1% p-D$_2$ in solid p-H$_2$ over that of pure p-H$_2$ [9]; × - $\Delta C_{L,Ne}$ induced by the change in the phonon spectrum of the crystal due to introduction of heavy quase-isotopic Ne impurity to the lattice of p-H$_2$.

The temperature dependences of the excess heat capacities $\Delta C_{Ne}(T)=C-C_1$ taken in series 1 – 4 are shown in Fig. 2. The figure also shows the temperature dependences of the excess heat capacity of solid solution of 1% p-D$_2$, 0.25% Ne in p-H$_2$ in comparison with the heat capacity of pure p-H$_2$ (Series 1), the excess heat capacity $\Delta C_{p-D2}(T)$ of the solid 1% p-D$_2$ – p-H$_2$ solution in comparison with the heat capacity of pure p-H$_2$ and the increment in the lattice heat capacity - $\Delta C_{L,Ne}$. Note that at $T<2$ K the contribution of $\Delta C_{L,Ne}$ to $\Delta C_{Ne}$ is negligible (see Fig.2). Therefore, the excess heat capacity $\Delta C_{Ne}$ is practically determined by the rotational motion of the p-D$_2$ molecules in the (p-D$_2$)Ne – type clusters. The temperature dependences $C_{R,Ne}(T)$ and $C_{R,p-D2}(T)$ [10] ($C_{R,p-D2}$ is the heat capacity of the rotational subsystem of the 1% p-D$_2$ – p-H$_2$ solution) are shown in Fig.3.

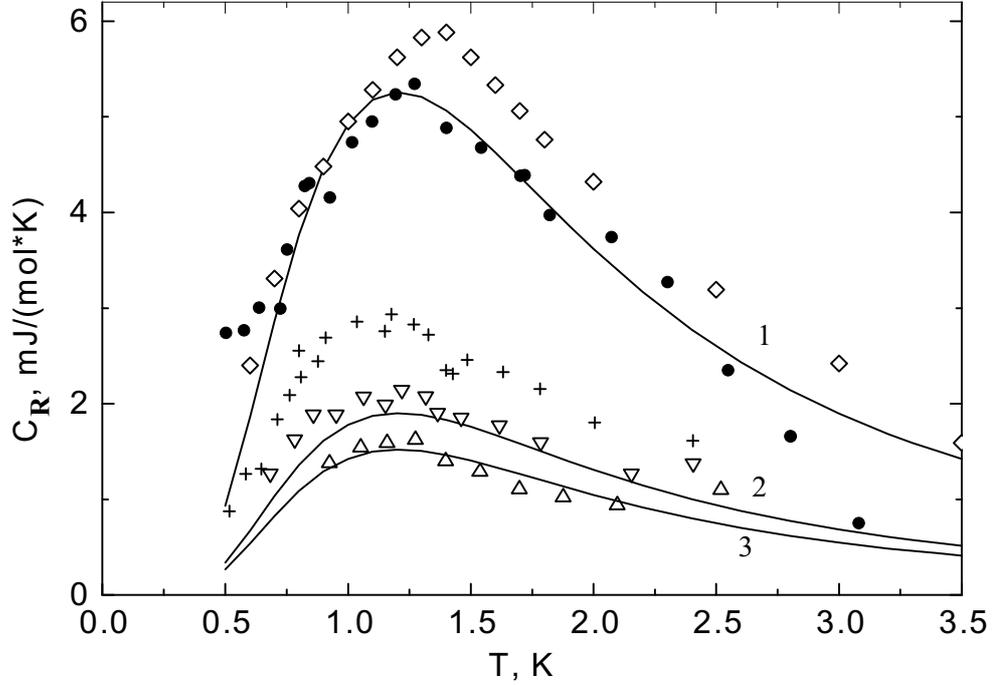

Fig. 3. Temperature dependences of heat capacities determined by rotation of p-D$_2$ molecules. Experiment: ●, +, △, ▽ - in (p-D$_2$)Ne clusters, solid solution 1% p-D$_2$, 0.25% Ne in p-H$_2$, series 1, 2, 3, 4 respectively; ◇ - in p-D$_2$p-D$_2$ clusters, solid solution 1% p-D$_2$ in p-H$_2$ [10]. The curves show calculated heat capacities $C_{R,Ne}$ for different contents of (p-D$_2$)Ne clusters: curve 2 – the number of clusters is $N_R$ at random distribution of Ne and p-D$_2$ impurities; curve 1 - the number of clusters is 2.8 times larger than $N_R$; curve 3 – the number of clusters is 1.25 times smaller than $N_R$.

The excess heat capacity $C_{R,Ne}$ was analyzed within the theoretical model of [2]. A number of new phenomena have been observed, which are induced by doping the solid 1% p-D$_2$ in p-H$_2$ solution with 0.25% Ne.

- An anomalously high excess heat capacity $\Delta C_{Ne}$ has been observed after

addition of 0.25% Ne to the solid 1% p-D$_2$ – p-H$_2$ solution. It is found that below 2 K the dominant contribution ($C_{R,\,Ne}$) to the heat capacity $\Delta C_{Ne}$ is made by the rotation of the p-D$_2$ molecules in the (p-D$_2$)Ne clusters (Fig.2).

- The heat capacity $\Delta C_{Ne}$ is strongly dependent on the method of preparation of a solid sample. Note that the excess heat capacity of the solid p-H$_2$ - 1% p-D$_2$ solution over that of pure p-H$_2$ is independent of the method of solid sample preparation.

- At $T$ < 3 K the temperature dependence of the excess heat capacity $C_{R,\,Ne}$ has the form of the Schottky curve and is described be the theory [2]. The splitting $\Delta$ = 3.2 ±0.1 K of the $J$ = 1 level of the p-D$_2$ molecules in the (p - D$_2$)Ne - type clusters was obtained from the analysis of $C_{R,\,Ne}(T)$ and is consistent with the theoretical estimate [2]. The number of (p-D$_2$)Ne clusters in the samples measured in Series 1 and 4 is 2.8 times larger and 1.25 times smaller than that in the case of randomly distributed p-D$_2$ and Ne impurities (see Fig. 3).

The effects observed evidence in favor of the existence of new condensable systems formed by the Van der Waals complexes of the Ne(H$_2$)$_n$ or Ne(D$_2$)$_n$ type [6, 7]. It has been found [3-7] that mixtures of quantum substances (e. g., helium and hydrogen) with inert elements or simple molecular substances can form Van der Waals complexes which make a basis for a new type of solids. X-ray investigations of Ne – containing H$_2$ and D$_2$ polycrystals samples prepared by condensation of gas mixtures on to a substrate at $T\approx$5 K show that in addition to the hexagonal and cubic phases based on the H$_2$ and Ne lattices, the samples contain hcp inclusion (even of the 0.25% Ne concentration [6, 7]) whose lattices have somewhat larger (by 1.5%-0.7%) volumes than that of pure Ne. The authors believe that the additional hcp phase in these systems is formed on the basis of the Ne(H$_2$)$_n$ or Ne(D$_2$)$_n$ types of Van der Waals complexes. We can assume that in the 1% p-D$_2$ and 0.25% Ne in p-H$_2$ solution the solid p-D$_2$ concentration produced by the Ne(p-H$_2$)$_n$ – type complexes and the amount of this phase are strongly dependent on the preparation conditions. This is because the formation of the (p-D$_2$)Ne clusters decreases the energy of the system by 2$\Delta$/3 and, hence, the total (elastic) energy of dilatation. In a liquid sample, the phase formed by the Ne(p-H$_2$)$_n$ complexes dissociates rather slowly, which reduces the number of (p-D$_2$)Ne clusters.

AKNOWLEDGMENTS

The authors are indebted to A.I. Prokhvatilov, and M.A. Strzhemechny for helpful discussions. The study was supported by the Ukraine Minister of Education and Science State Foundation of for Basic research (Project № 02.07/00391-2004).